\documentstyle[12pt]{article} 

\def\gsim{\:\raisebox{-0.5ex}{$\stackrel{\textstyle>}{\sim}$}\:}

\begin{document}
\begin{flushright} TUM-HEP-533/03 \\
October 2003
\end{flushright}

\vspace*{1cm}
\begin{center}

{\Large \bf The Top--Down Interpretation of Ultra--High Energy Cosmic
Rays}\footnote{Invited talk at the {\it 3rd International Workshop for 
Comprehensive Study of the High Energy Universe - Toward Very High
Energy Particle Astronomy - (VHEPA-3)}, ICRR, Tokyo, March 2003.} \\
\vspace*{6mm}
\textsc{Manuel Drees} \\
\vspace*{3mm}
{\it Physik Department, TU M\"unchen, D--85748 Garching, Germany}

\end{center}
\vspace*{1cm}

\begin{abstract}
The origin of Ultra--High Energy ($E \, \gsim\, 10^{20}$ eV) Cosmic
Rays (UHECR) remains mysterious. I discuss ``top--down'' models, where
UHECR originate from the decay of very massive, long--lived
particles. I summarize the calculation of the spectrum of decay
products, discuss possible problems with this scenario, and describe
ways to test it by searching for very energetic neutrinos and neutralinos.
\end{abstract}

\vspace*{1cm}

\noindent
The observation \cite{md1} of CR events with $E \,\gsim\, 10^{20}$ EeV
poses two problems:
\begin{itemize}
\item {\bf The energy problem:} Few, if any, known celestial objects
possess sufficiently strong electromagnetic fields extended over a
sufficiently large length to accelerate protons to such energies.

\item {\bf The propagation problem:} Protons with $E \geq 5 \cdot
10^{19}$ eV lose energy by photoproducing pions on CMB photons
\cite{md4}. This should lead to a sharp drop in the spectrum if
sources are distributed homogeneously (the ``GZK cut--off'').
Similarly, heavier nuclei lose energy through photo--dissociation, and
photons get converted into $e^+e^-$ pairs \cite{md5}.
\end{itemize}
The second problem aggravates the first one: The GZK effect implies
that sources of UHECR should be within $\sim 50$ Mpc. Even charged
particles (in particular, protons) should travel along rather straight
lines over such distances, once $E > 10^{19}$ eV. However, there
are no known nearby sources in the UHECR arrival directions.

In top--down models \cite{md7} the energy problem is solved by
postulating the existence of ``$X$'' particles with mass $m_X \geq
10^{12}$ GeV and lifetime $\tau_X \geq 10^{10}$ yrs $\simeq
\tau_U$. The decay of such objects clearly can generate particles of
the required energy. Moreover, several mechanisms for producing a
sufficient abundance of $X$ particles in the very early Universe have
been proposed \cite{md8}. Finally, $X$ particles can be very
long--lived either by embedding them in topological defects
\cite{md7,md5}, or by greatly suppressing their couplings to lighter
particles \cite{md9}. Essentially $X$ particles are batteries storing
energy from an earlier, much more violent epoch until the present
time.

The propagation problem can also be solved in this scenario. If $X$
particles are distributed more or less evenly throughout the Universe,
as expected e.g. in topological defect models \cite{md5}, a spectral
break at the GZK energy is expected, in agreement with the HiRes (but
not the AGASA) data, but there might be some ``nearby'' sources not
associated with known celestial objects. Recall that HiRes does see
some post--GZK events, which implies the existence of ``nearby''
sources (unless Lorentz invariance is violated \cite{md11}.)
Alternatively, if $X$ particles move freely under the influence of
gravity, an enhancement in their density by a factor $\sim 10^5$ is
expected in the halo of our own galaxy. In this case almost all
observed UHECR events would be of ``local'' origin, and no sharp
drop--off would be expected at the GZK energy, in agreement with the
AGASA (but not the HiRes) data \cite{md1}. In both cases the spectrum
should cut off at $E = m_X/2$.

In order to test this idea quantitatively, one first has to compute
the spectrum of $X$ decay products. The very existence of an energy
scale $m_X$ strongly indicates the existence of weak--scale
supersymmetry (SUSY), in order to stabilize the hierarchy between
$m_X$ and the weak scale ($\sim 100$ GeV) against radiative
corrections \cite{md13}. One then expects primary $X$ decays to
produce comparable numbers of ordinary particles and their
superpartners. This primary $X$ decay will start a parton cascade,
similar to $Z \rightarrow q \bar q$ decays at LEP starting a QCD
cascade leading to a multi--parton state. There are a couple of
important differences, though, related to the fact that the particles
produced in primary $X$ decay will have time--like virtuality near
$m_X$. First, at such scales all three gauge interactions have
comparable strengths, and should therefore be treated on equal
footing.  Secondly, at such scales superparticles can be treated as
massless, and can thus be produced during the cascade, along with
quarks, gluons, leptons etc.

Once the cascade virtuality scale drops below the mass scale of SUSY
particles $m_{\rm SUSY} \leq 1$ TeV, superparticles (as well as top
quarks and electroweak gauge and Higgs bosons) decouple from the
cascade and decay (often starting their own little QCD cascades). At
virtuality $< m_{\rm SUSY}$ only strong interactions need to be kept
in the description. Finally the virtuality scale reaches $Q_{\rm had}
\simeq 1$ GeV, where partons hadronize into baryons and mesons, many
of which in turn decay (along with the heavy leptons).

This cascade started by $X$ decays has been treated in \cite{md14}
using the language of fragmentation functions, which allow relatively
easy calculation of the single--particle inclusive
spectra\footnote{Our code computing the $X$ decay spectrum is publicly
available \cite{md15}}, exactly the quantities needed in our case
(since at most one particle produced in a given $X$ decay will be
detected on Earth). Not surprisingly, the high--energy end of the
resulting spectrum depends strongly on the most important primary $X$
decay mode(s). If $X$ decays into ordinary particles, most of the
energy will be carried away by neutrinos and photons; protons carry at
most 10\%. If the primary decay products are superparticles, LSPs
often carry the largest fraction of the released energy; hundreds of
LSPs are produced even in $X \rightarrow q \bar q$ decays. On the
other hand, the shape of the proton, photon, electron, and electron
and muon (anti)neutrino spectra at energies $\leq 0.01 m_X$ are
essentially independent of the primary $X$ modes, and depend only
logarithmically on $m_X$ if they are expressed in terms of the scaling
variable $x = 2 E_{\rm particle} / m_X$.

Top--down models face several challenges:
\begin{itemize}

\item
For all primary $X$ decay modes that have been investigated [including
the whole catalogue of (s)particles that exist in the MSSM] the
$\gamma$ flux is significantly higher than the proton flux {\em at
source}. This is problematic, since various independent investigations
of Haverah Park, Fly's Eye and AGASA data indicate that relatively
few, if any, UHECR events are due to photons \cite{md16}. If most
sources are at cosmological distances, propagation effects increase
the $p/\gamma$ ratio. For sources in the halo of our galaxy
propagation should be significant only if the galactic radio
background has been underestimated by at least a factor of
10. Moreover, absorbing UHE photons has its own problem: It initiates
an electromagnetic cascade, which eventually produces photons in the
100 MeV to 1 TeV energy range probed by the EGRET satellite. The upper
bound on the isotropic flux of such photons imposes strong constraints
on top--down models \cite{md17}. This is probably the most serious
current problem for top--down models. Possible excuses are the rather
poor statistics at very high energy, the fact that constraining the
observed $p/\gamma$ ratio heavily relies on MC studies, or the fact
that the crucial EGRET bound has not been checked independently.

\item
Precision measurements of the CMB anisotropy \cite{md18} indicate that
topological defects play little or no role in structure formation in
the Universe. It is not easy to square this with the requirement that
the (symmetry breaking) scale associated with the defects has to
exceed $10^{12}$ GeV in order to explain the UHECR. Freely moving
particles thus seem the more likely option. As noted earlier, they
should be concentrated in the halo of our galaxy. Since we are not in
the center of our galaxy, one expects an anisotropy in the UHECR
arrival directions in this ``decaying dark matter'' (DDM) model
\cite{md19}, in particular a peak roughly in the direction of the
galactic center, where the DM density should be highest. This location
is not visible from the Northern hemisphere, but could be observed by
the Australian SUGAR array. The data from this experiment show no such
peak \cite{md20}. However, the size and exact location of the expected
peak depends on details of the DM distribution near the galactic
center, which is poorly understood, and also on the extension of the
dark halo of our galaxy, which is unknown. Even taking standard halo
models at face value, SUGRA data are compatible at the 10\% level with
all events above $6 \cdot 10^{19}$ eV being due to DDM. On the other
hand, if the Auger array in Argentina also fails to see any
enhancement towards the galactic center, the DDM variety of top--down
models would be in serious difficulties.

\item
The AGASA data above $10^{19}$ eV show quite pronounced (at the 4 to
5$\sigma$ level) clustering on small scales \cite{md21}. The HiRes
collaboration is currently not claiming such clustering, but also does
not seem to be willing to state (yet another) discrepancy with
AGASA. Although some small--scale anisotropy can be expected in DDM
models with clumpy halo, the clumpiness required to explain the AGASA
data does not seem to be compatible with current halo models
\cite{md22}. Note that here the structure of the halo in our
``vicinity'' is most important, which is supposed to be fairly well
understood. However, even in the AGASA data the evidence for
clustering becomes weak if one restricts oneself to events (well)
above the GZK cut--off; this is opposite to what one expects in
bottom--up models, where the number of relevant sources should
decrease rapidly beyond the GZK cut--off. In any case, the present
data on small--scale clustering do not exclude all UHECR with $E \,
\gsim \, 10^{20}$ eV being of top--down origin. Similar remarks apply
for the claimed correlation of UHECR with $E \sim$(a few) times
$10^{19}$ eV with certain BL Lac's \cite{md23}.

\item
The last two points indicate that many, perhaps even most, events
right at and below the GZK cut--off should {\em not} be of top--down
origin, whereas the energy problem indicates that this should be the
dominant source at $E \geq 10^{20}$ eV. Shouldn't one expect some sort
of spectral feature in the transition region? Generally speaking, Yes,
but this argument is difficult to make more quantitative as long as
the origin of CR's with $E \sim 10^{19}$ eV is not understood in
detail. Note also that the AGASA data {\em are} compatible with a
spectral break (a hardening of the spectrum) in this energy range.

\end{itemize}

It should be noted that top--down models are more vulnerable than
``generic'' bottom--up models because they make detailed predictions
for the spectrum at source (for given $X$ decay mode), and in case of
DDM models also about the distribution of sources. In contrast,
bottom--up ``models'' often simply assume that protons are injected
with a given spectrum (usually a power law with a cut--off), without
bothering to explain {\em how} the protons were accelerated. Moreover,
often a continuous distribution of sources is assumed (e.g. in the
``model'' adapted by HiRes to explain their data \cite{md1}), even
though the number of possible bottom--up sources within a few GZK
interaction lengths is certain to be quite small. To my mind this is
only a fancy parameterization of the data, not a physical model.

As discussed above, the more predictive, and currently favored, DDM
variety of top--down models makes predictions that can be tested by
Auger. Other tests of top--down models are based on predictions for
neutrino\cite{md24a} and neutralino (LSP) fluxes. These signals would
be even bigger (by a factor $\sim 10$) if most sources were at
cosmological distances, since one would then require a larger flux at
source to explain the UHECR data. In DDM scenarios, normalizing the
observed UHECR flux to the proton flux only, one expects the following
signals:\footnote{Rates are typically three times smaller if the sum
of photon and proton fluxes is normalized to the UHECR flux.}

\begin{itemize}
\item {\bf Neutrinos} \cite{md24}: 1--30 (mostly $\nu_\mu$)
events$/$yr above 100 TeV in IceCube, 0.4--4 (mostly $\nu_e$)
events$/$yr in RICE, and 0.3--3 (mostly $\nu_\tau$) events$/$yr in
Auger. The ranges correspond to variations in $m_X$ and the primary
$X$ decay mode, with smaller $m_X$ and less hadronic $X$ decays giving
higher rates. However, less hadronic decays also give a larger
$\gamma/p$ ratio at source, and hence a bigger problem with either the
data \cite{md16} favoring protons as UHECR primaries, or with the
EGRET bound \cite{md17}. However, most other UHECR models also lead to
detectable neutrino fluxes at $E > 100$ TeV, where the atmospheric
background becomes negligible \cite{md17}. The neutrino flux should
thus give valuable quantitative tests, but does not distinguish models
qualitatively.

\item {\bf Neutralinos} \cite{md25}: Only top--down models can produce
a detectable flux of UHE neutralinos, since only here the flux of
particles and sparticles at source are comparable. Note that LSPs are
expected to have 10--100 times smaller interaction cross section with
matter than neutrinos \cite{md26}. There thus exists a range of
energies (between a few $10^8$ GeV and a few $10^{10}$ GeV) where
neutralinos can still cross the Earth more or less unmolested whereas
neutrinos get stuck. Looking for UHE LSPs thus boils down to looking
for clearly upgoing events in this energy window. Unfortunately the
expected event rate is marginal even for (first--generation) space
based experiments like EUSO, but at least in principle this signal is
detectable.

\end{itemize}

In summary, CR events with $E \, \gsim\, 10^{20}$ eV may well
originate from the decay of very massive particles; however, this
contribution to the CR spectrum should be sub--dominant for $E \leq 5
\cdot 10^{19}$ eV. There may therefore exist two distinct UHECR
puzzles. The component(s) at $E \leq 5 \cdot 10^{19}$ eV would have to
come from some acceleration mechanism. Crucial tests of this picture
should come from better measurements of the UHECR spectrum, and from
analyses of the clustering of UHECR, and/or their correlation with
known celestial objects (e.g. the center of the Milky Way). A complete
understanding will probably require studies of the UHECR sources using
penetrating particles. In this picture, both bottom--up and top--down
sources are likely to contribute to the spectrum of neutrinos at
$E_\nu > 100$ TeV; a detailed measurement of the spectrum of these
neutrinos, while difficult, should then be rewarding. Detecting UHE
neutralinos would constitute a ``smoking gun'' signature for top--down
models.

\section*{Acknowledgment}
I thank my collaborators Cyrille Barbot, Francis Halzen and Dan
Hooper for the collaborations that allowed me to attend this
interesting workshop. This work was supported by the SFB 375 of the
Deutsche Forschungsgemeinschaft.


\begin{thebibliography}{99} %% The number "99" means that this list
			    %% has more than nine items. 

\bibitem{md1}
M. Nagano and A.A. Watson, Rev. Mod. Phys. {\bf 72}, 689 (2000). For
more recent data, see
HiRes-MIA collab., T.~Abu-Zayyad et al., astro--ph/0208243;
AGASA collab., M. Takeda et al., Astropart. Phys. {\bf 19}, 447 (2003).

\bibitem{md4}
K.~Greisen, Phys. Rev. Lett. {\bf 16}, 748 (1966); G.T.~Zatsepin
and V.A.~Kuzmin, JETP Lett. {\bf 4}, 78 (1966) [Pisma Zh. Eksp.
Teor. Fiz. {\bf 4}, 114 (1966)].

\bibitem{md5}
For a review, see P.~Bhattacharjee and G.~Sigl, Phys. Rept. {\bf 327},
109 (2000). 

\bibitem{md7}
C.T. Hill, D.N. Schramm and T.P. Walker, Phys. Rev. {\bf D36}, 1007
(1987);
P. Bhattacharjee, C.T. Hill and D.N. Schramm, Phys. Rev. Lett. {\bf
69}, 567 (1992).

\bibitem{md8}
D.J.H. Chung, E.W. Kolb and A. Riotto, Phys. Rev. {\bf D59}, 023501
(1999); 
D.J.H. Chung, P. Crotty, E.W. Kolb and A. Riotto, Phys. Rev. {\bf D64}
043503 (2001); 
R. Allahverdi and M. Drees, Phys. Rev. Lett. {\bf 89}, 091302 (2002).

\bibitem{md9}
J. Ellis, J. Lopez and D.V. Nanopoulos, Phys. Lett. {\bf B247}, 257
(1990);
S. Chang, C. Coriano and A.E. Faraggi, Nucl. Phys. {\bf B477}, 65
(1996); 
K. Benakli, J.R. Ellis and D.V. Nanopoulos, Phys. Rev. {\bf
D59}, 047301 (1999).

\bibitem{md11}
H. Sato and T. Tati, Prog. Theor. Phys. {\bf 47}, 1788 (1972).

\bibitem{md13}
For a review, see H.--P. Nilles, Phys. Rep. {\bf 110}, 1 (1984).

\bibitem{md14}
C. Barbot and M. Drees, Phys. Lett. {\bf B533}, 107 (2002), and
Astropart. Phys. {\bf 20}, 5 (2003). For earlier, less complete,
studies see e.g. M.~Birkel and S.~Sarkar, Astropart. Phys. {\bf 9}, 297 (1998);
V. Berezinsky and M. Kachelriess, Phys. Rev. {\bf D63}, 034007 (2001);
%Z. Fodor and S.D. Katz, Phys. Rev. Lett. {\bf 86}, 3224 (2001);
S.~Sarkar and R.~Toldra, Nucl.\ Phys.\ {\bf B621}, 495 (2002).

\bibitem{md15}
C. Barbot, hep--ph/0306303, to appear in Comput. Phys. Commun. 

\bibitem{md16}
R.A. Vazquez et al., Astropart. Phys. {\bf 3}, 151 (1995); 
M.~Ave et al., Phys. Rev. {\bf D65}, 063007 (2002); 
K. Shinozaki et al., AGASA collab., Astrophys. J. {\bf 571}, L120 (2002).

\bibitem{md17}
See e.g. D.V. Semikoz and G. Sigl, hep--ph/0309328, and references therein.

\bibitem{md18} 
WMAP collab., C.L. Bennett et al., Astrophys. J. Suppl. {\bf 148}, 1
(2003).

\bibitem{md19}
S.L. Dubovsky and P.G. Tinyakov, JETP Lett. {\bf 68}, 107 (1998);
V. Berezinsky, P. Blasi and A. Vilenkin, Phys. Rev. {\bf D58}, 103515
(1998);
N.W. Evans, F. Ferrer and S. Sarkar, Astropart. Phys. {\bf 17}, 319
(2002). 

\bibitem{md20}
H.B. Kim and P. Tinyakov, astro--ph/0306413;
M. Kachelriess and D.V. Semikoz, astro--ph/0306282.

\bibitem{md21}
M. Takeda et al., AGASA collab., astro--ph/9902239;
W.S. Burgett and M.R. O'Malley, Phys. Rev. {\bf D67}, 092002 (2003).

\bibitem{md22}
N.W. Evans, F. Ferrer and S. Sarkar, Phys. Rev. {\bf D67}, 103005
(2003). 

\bibitem{md23}
P.G. Tinyakov and I.I. Tkachev, JETP Lett. {\bf 74}, 445 (2001), and
astro--ph/0301336. 

\bibitem{md24a}
P. Gondolo, G. Gelmini and S. Sarkar, Nucl. Phys. B {\bf 392}, 111
(1993).

\bibitem{md24}
C. Barbot, M. Drees, F. Halzen and D. Hooper, Phys. Lett. {\bf B555},
22 (2003).

\bibitem{md25}
C. Barbot, M. Drees, F. Halzen and D. Hooper, Phys. Lett. {\bf B563},
132 (2003).

\bibitem{md26}
V. Berezinsky and M. Kachelriess, Phys. Lett. {\bf B422}, 163 (1998).

\end{thebibliography}
\end{document}